\documentclass[iop,apj,twocolumn]{emulateapj_pab}
\RequirePackage{hyperref}
\hypersetup{
	colorlinks = true,
	linkcolor = blue,
	citecolor = blue,
	filecolor = blue,
	urlcolor = blue,
	pdfborder = {0 0 0}
}

\usepackage{color}
\usepackage{bm}

\renewcommand{\vec}[1]{\bm{#1}}

\graphicspath{{./}{./Plots/}}

\renewcommand*\aap{A\&A}

\renewcommand*\ao{Appl.~Opt.}

\renewcommand*\apjl{ApJL}

\renewcommand*\apjs{ApJS}

\received{2019 March 28}
\accepted{2019 April 29}
\published{2019 June 11}

\submitted{}
\journalinfo{The Astrophysical Journal, {\rm 878:30 (6pp), 2019 June 10 \hfill \url{https://doi.org/10.3847/1538-4357/ab1e48}}}
\shorttitle{Electromotive force as a shock front indicator}
\shortauthors{Hofer \& Bourdin}

\begin{document}

\title{
Application of the Electromotive Force as a Shock Front Indicator in the Inner Heliosphere
}

\author{Bernhard Hofer$^{1,2}$ \orcid{0000-0002-8628-7887}, Philippe-A.~Bourdin$^{1}$ \orcid{0000-0002-6793-601X}}
\affiliation{$^{1}$Space Research Institute, Austrian Academy of Sciences, Schmiedlstr. 6, A-8042 Graz, Austria}
\affiliation{$^{2}$Max-Planck-Institut f{\"u}r Sonnensystemforschung, Justus-von-Liebig-Weg 3, D-37077 G{\"o}ttingen, Germany, \href{mailto:hofer@mps.mpg.de}{hofer@mps.mpg.de}}


\begin{abstract}

The electromotive force (EMF) describes how the evolution and generation of a large-scale magnetic field is influenced by small-scale turbulence. 
Recent studies of in situ measurements have shown a significant peak in the EMF while a coronal mass ejection (CME) shock front passes by the spacecraft.
The goal of this study is to use the EMF as an indicator for the arrival of CME shock fronts.
With {\em Helios} spacecraft measurements we carry out a statistical study on the EMF during CMEs in the inner heliosphere.
We develop an automated shock front detection algorithm using the EMF as the main detection criterion and compare the results to an existing CME database.
The properties of the EMF during the recorded events are discussed as a function of the heliocentric distance.
Our algorithm reproduces most of the events from \cite{Kilpua+al:2015} and finds many additional CME-like events, which proves that the EMF is a good shock front indicator.
The largest peaks in the EMF are found from 0 to 50 minutes after the initial shock.
We find a power law of $-1.54$ and $-2.18$ for two different formulations of the EMF with the heliocentric distance.

\end{abstract}



\section{Introduction} \label{introduction}

The study of small-scale turbulence and its physical effects within a plasma is a scientific field of great interest \citep{Bruno-Carbone:2013}. 
For instance, in turbulent dynamo theory, the turbulent electromotive force (EMF) can give rise to a large-scale magnetic field through mutual interactions of small-scale plasma motions and small-scale magnetic fields; see \cite{Brandenburg-Subramanian:2005} and works cited therein. 
Research on the EMF is conducted for various purposes, one particularly prominent research area is nuclear fusion \citep{Ji-Prager:2002} and studies under laboratory conditions.
Direct numerical simulations (DNS) have become increasingly popular with advances in computer technology.
DNSs have the advantage that physical effects can be investigated in great detail and under conditions that can currently not be reached by technical means.
This allows the study of distinct magnetic field generation mechanisms such as the $\alpha$ effect \citep{Brandenburg-Subramanian:2005b} and the cross-helicity effect \citep{Yokoi:2013}.
The disadvantages of DNSs are other limitations such as the necessity of artificial boundary conditions.

In comparison, the number of investigations of the EMF in astrophysical plasmas by in situ measurements, such as in the solar wind, is very limited. 
The pioneering work of \cite{Marsch-Tu:1992} used {\em Helios} spacecraft measurements of the proton bulk velocity and the magnetic field to calculate the EMF and the $\alpha$ effect within the inner heliosphere. 
As expected, the $\alpha$ effect and EMF are found to be negligible and therefore there is no relevant dynamo action within the solar wind.
However, \cite{Narita-Vörös:2018} and \cite{Bourdin+al:2018}, also using {\em Helios} spacecraft data, find that during coronal mass ejections (CMEs) there can be significant peaks in the EMF when compared to the quiet solar wind. 
\cite{Bourdin+al:2018} further reproduce the signatures of the EMF and turbulent transport coefficients with a simplified shock front model and suggest the possibility of using the EMF as an {\em in situ} indicator for the arrival of shock fronts. 

The goal of this study is to investigate the relationship between the EMF and CMEs in the inner heliosphere. 
We use the EMF as a detection criterion for shock fronts of magnetic transient events passing by the spacecraft.
Additionally, we develop a numerical algorithm to detect shock fronts from in situ measurements of the plasma flow velocity $U$, the magnetic field $B$, the proton number density $n_P$, and the proton temperature $T_P$.
In order to analyze the EMF at different heliocentric distances we use measurements of the {\em Helios} spacecraft as their orbits range between 0.3 and 1 astronomical units (au) from the Sun.
In this work we list magnetic transient events that feature shock fronts and significant peaks in the EMF and compare them to an existing CME database.
Finally, we analyze and calculate a power law of the amplitude of the EMF versus the heliocentric distance during the ensemble of events.

\subsection{Electromotive Force}

\subsubsection{Mean-field Electrodynamics}
The EMF can be directly derived from mean-field electrodynamics; see \cite{Steenbeck+al:1966} and \cite{Krause-Raedler:1980}.
Here, the magnetic field $\vec{ B}$ and plasma flow velocity $\vec{ U}$ are separated into large-scale background fields ($\vec{ B}_0$, $\vec{ U}_0$) and small-scale fluctuation fields ($\delta \vec{ B}$, $\delta \vec{ U}$).
We obtain the EMF $\vec{ M}_1$ from the mean induction equation:

\begin{equation}
\frac{\partial \vec{ B}_0}{\partial t} = \vec{ \nabla} \times (\vec{ U}_0 \times \vec{ B}_0) + \vec{ \nabla} \times \vec{ M}_1 + \eta \vec{ \nabla}^{2} \vec{ B}_0
\label{eq:ind}
\end{equation}
\\
In this equation, the EMF describes the impact of small-scale fluctuations on the large-scale magnetic field:

\begin{equation}
\vec{ M}_1 =  \langle \delta \vec{ U} \times \delta \vec{ B} \rangle,
\label{eq:em}
\end{equation}
\\
where $\langle$ $\rangle$ denotes the ensemble average.

\pagebreak

\subsubsection{Reversed Field Pinch}
\cite{Bourdin+al:2018} also use a second formulation of the EMF $\vec{ M}_2$ that was adapted from the reversed field pinch \citep[RFP; see][]{Bodin-Newton:1980} model of \cite{Yoshizawa:1990} to the application of in situ measurements in the solar wind:
\begin{equation}
\vec{ M}_2 = \alpha \vec{ B}_0 - \beta (\vec{ \nabla} \times \delta \vec{ B}) + \gamma (\vec{ \nabla} \times \delta \vec{ U}), 
\label{eq:em2}
\end{equation}
where $\alpha$, $\beta$ and $\gamma$ are so-called turbulent transport coefficients that correspond to different physical effects within a plasma.
The first ($\alpha$) and third ($\gamma$) terms in Eq.~(\ref{eq:em2}) describe magnetic field generation mechanisms, the kinetic helicity effect, and the cross-helicity effect.
The second ($\beta$) term in Eq.~(\ref{eq:em2}) corresponds to magnetic diffusion of the large-scale magnetic field.
In \cite{Bourdin+al:2018} the turbulent transport coefficients were chosen according to \cite{Krause-Raedler:1980} as:
\begin{eqnarray}
\alpha &=& \textstyle{\frac{1}{3}} \tau \langle -\delta \vec{ U} \cdot (\vec{ \nabla} \times \delta \vec{ U}) \rangle \label{eq:alpha1}
\\
\beta &=& \textstyle{\frac{1}{3}} \tau \langle \delta \vec{ U} \cdot \delta \vec{ U} \rangle, \label{eq:beta1}
\\
\gamma &=& \textstyle{\frac{1}{3}} \tau \langle \delta \vec{ U} \cdot \delta \vec{ B} \rangle. \label{eq:gamma1}
\end{eqnarray}
where $\tau$ is a characteristic time scale that describes the decay of turbulent energy and helical structures in the solar wind; see also \cite{Bourdin+al:2018}.
\\

\subsection{Interplanetary Coronal Mass Ejections (ICMEs)} \label{ICME}

CMEs are massive outbursts of plasma and magnetic fields that originate from instable active regions on the solar surface.
During a complex trigger mechanism that includes magnetic reconnection, CMEs can be accelerated to high velocities ranging from several hundreds of km s$^{-1}$ to more than 1000 km s$^{-1}$; see \cite{Chen:2011} and references therein.
When a CME propagates into interplanetary space it is commonly referred to as an ICME.

\subsubsection{In Situ Signatures of ICMEs} \label{in-situ}
In situ spacecraft measurements give information on the internal structure of ICMEs that would not be available from remote-sensing observations alone.
When an ICME passes by the spacecraft at a propagation velocity that exceeds the local magnetosonic speed, one observes a leading shock front that can be measured as a sudden increase in velocity followed by a highly turbulent "sheath region" that has an enhanced magnetic field magnitude; see \cite{Kilpua+al:2017}.
Due to compression one also finds enhanced plasma density and temperature.
The sheath region is followed by the driving ICME structure that may also contain a magnetic cloud or a flux rope structure that is characterized by a helical magnetic field; see \cite{Burlaga+al:1981} and \cite{Klein-Burlaga:1982}.
ICMEs can be identified by a large number of characteristic signatures such as enhanced magnetic field, reduced proton temperature, or increased $\alpha$-to-proton ratio.
However, there are many ICMEs that lack some of the characteristics that make an unambiguous definition difficult \citep{Zurbuchen-Richardson:2006}.
Therefore, there are always differences between various ICME lists depending on the applied detection criteria.  

We introduce an automated shock front detection method in Sect.~\ref{methods}, based on EMF measurements and a numerical detection algorithm.
After choosing the detection parameters in Sect.~\ref{parameters}, the method is applied to the {\em Helios} spacecraft data.
We create a list of magnetic transient events and calculate the average detection rate with the heliocentric distance; see Sect.~\ref{list}.
Additionally, we fit a power law to the magnitude of the EMF versus the heliocentric distance of the recorded transient events; see Sect.~\ref{power}.
Finally, in Sect.~\ref{ipshocks}, we compare our results to an existing database and analyze the time differences between the EMF peaks and the initial shock front.


\section{Methods} \label{methods}

We use magnetic field and plasma measurements from the {\em Helios} spacecraft to create a list of magnetic transient events in the inner heliosphere.
Following the results of~\cite{Bourdin+al:2018} we apply the EMF as a detection criterion for the arrival of shock fronts.
We further develop a numerical algorithm to automatically detect shock fronts from in situ measurements of the magnitude of the plasma flow velocity $U$, the magnetic field $B$, the proton number density $n_P$ and the proton temperature $T_P$.

\subsection{Mean-field Determination} \label{smooth}

As a first step, we need to separate the magnetic field and plasma velocity data into mean field and fluctuation.
The mean fields are calculated as a moving average that uses a Gaussian kernel instead of a rectangular boxcar function to avoid unwanted edges and we obtain the fluctuations as the residuals. 
We set the standard deviation $\sigma_0$ of this Gaussian-convolution filter as 11 data points at a time resolution of $\Delta t = 648$~s, identical to \cite{Bourdin+al:2018}.

\subsection{Shock-front Detection Algorithm} \label{algorithm}

When trying to detect shock fronts from the time series data, one needs to avoid false detections at fluctuation peaks.
The fluctuation peaks can reach similar amplitudes to the shock fronts and occur equally abruptly, so a simple numerical derivative will lead to false detections.
After a real shock front the magnitude of the measured quantity remains enhanced also after the shock front has passed.
In contrast, fluctuation peaks appear only on short time scales.
The moving median difference is robust to outliers and can be used to detect enhancements on time scales longer than normal fluctuations. 
For that, we calculate the median of $k$ data points before and after a central value $P_n$ and calculate the difference of these two medians: 
\begin{equation}
\overline{P_n} = {\rm median}(P_{n-k},...,P_n) - {\rm median}(P_{n},...,P_{n+k})
\label{eq:median_smooth}
\end{equation}

\subsubsection{Idealized Signal} \label{idealized}
\begin{figure}[ht]
\begin{center}
\includegraphics[trim=0 0 0 0,width=8.0cm]{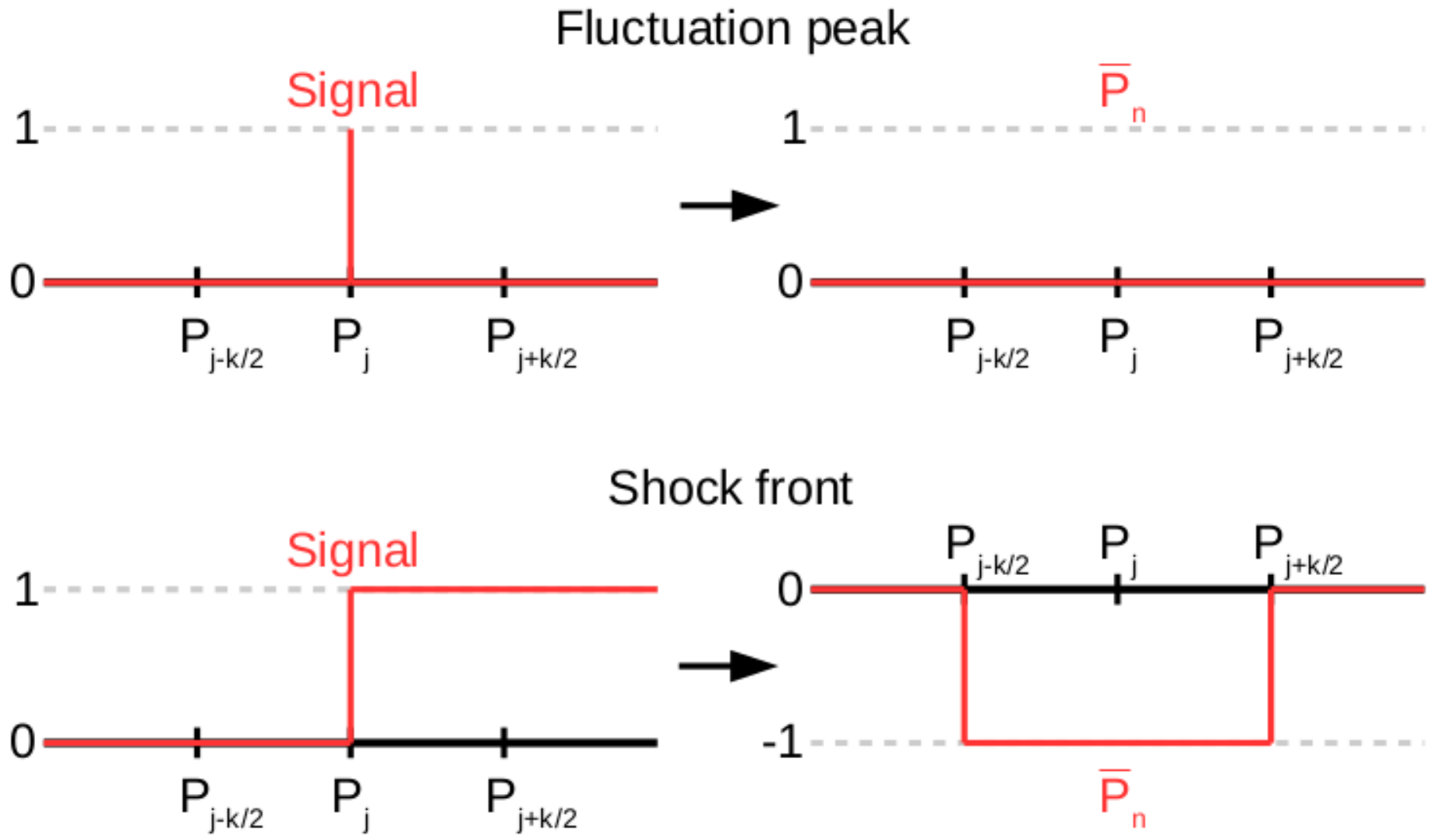}
\end{center}
\vspace*{-5mm}
\caption{Result of the moving median-difference algorithm -- see Eq.~(\ref{eq:median_smooth}) -- applied to an idealized signal.
Upper panel: the fluctuation peak (delta function) is ignored by the algorithm.
Lower panel: the shock front (step function) is detected by the algorithm.}
\label{fig:median_difference}
\end{figure}

\begin{figure}[ht]
\begin{center}
\includegraphics[trim=0 0 0 0,width=8.0cm]{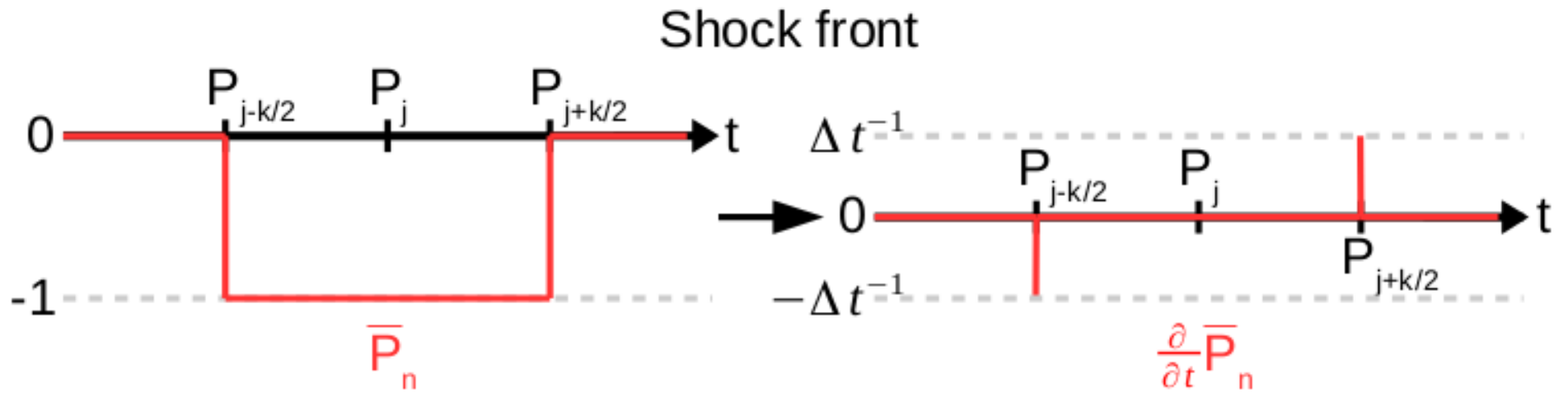}
\end{center}
\vspace*{-5mm}
\caption{Mirror peaks in the time derivative of the result of the moving median-difference algorithm -- see Eq.~(\ref{eq:median_smooth}) -- applied to an idealized signal.}
\label{fig:median_difference_2}
\end{figure}

Let us look at an idealized fluctuation peak and shock front to study the results we obtain for $\overline{P_n}$; see Fig.~\ref{fig:median_difference}.
We imagine the fluctuation peak is a single pulse that is $0$ everywhere but at the singular point $P_j$ with a value of $1$. 
Because the median is robust to outliers the median remains unchanged by such a fluctuation peak and $\overline{P_n} = 0$ everywhere.
In contrast, when there is a shock front, the measured quantity stays enhanced for some time, so imagine a step function with an initial value of $0$ that jumps onto the value of $1$ at the point $P_j$.
Initially $\overline{P_n}$ is again zero but when $P_n$ is $k/2$ points before $P_j$ the median of the points after will suddenly jump to $1$ and $\overline{P_n} = -1$.
This remains until $P_n$ is $k/2$ points past $P_j$ so that the median of the precedent point jumps to $1$ as well and the median difference is again $\overline{P_n} = 0$. 
When calculating the numerical derivative of $\overline{P_n}$, we find a peak $k/2$ points before the shock front and a peak with opposite sign $k/2$ points after; see Fig.~\ref{fig:median_difference_2}.
The center of the shock front at $P_j$ can be found as the central point between the mirror peaks at $P_{j-k/2}$ and $P_{j+k/2}$.  

\subsubsection{Application to In Situ Measurements} \label{application}

When applying the algorithm to in situ solar wind data, the results from the median analysis will not be as smooth as in the idealized signal case.
The background may not be constant and strong fluctuations may alter the median value as well. 
Implementing a detection method based on $P_n$ is difficult, because it requires testing a whole interval in the time series, while detecting mirror peaks in the time derivative $\frac{\partial}{\partial t}\overline{P_n}$ becomes simple with a minimum/maximum search.

The mirror peaks from the detected shock fronts in $\frac{\partial}{\partial t}\overline{P_n}$ are superimposed by fluctuations.
Also, the width of the moving median-difference $k$ has to be carefully chosen to be able to detect narrow shock fronts while avoiding the detection of fluctuations as a signal.
Before applying the algorithm, we perform a binning algorithm to ensure that the algorithm uses the median of an equally long time interval on both sides.
We then split the time series into $96.3$\,hr subintervals and transform each subinterval to streamwise coordinates.
There is an overlap of $6.3$\,hr between the intervals to avoid missing shock fronts at the boundary.
Possible double detections at the overlapping boundaries are filtered out afterwards.

To avoid false detections, we set a threshold $\chi$ that scales with the Sun-spacecraft distance and the sampling rate from the binning algorithm.
We measure the local maxima on the intervals of $\pm k/2$ points and check for a mirrored peak on the following interval.
Because of the fluctuations in $\frac{\partial}{\partial t}\overline{P_n}$ we apply a tolerance interval for the position of the second mirror peak of $\delta_{tol} = \pm 1$ points.
On successful detection, we set the center of the shock front $k/2$ points after the first mirror peak.
We search for shock fronts using $U$, $B$, $n_p$ and $T_p$ in the mirrored median way. 
Simultaneously, we compute the EMF $M_1$ and $M_2$ from Eqs.~(\ref{eq:em}) and~(\ref{eq:em2}).
We set an individual threshold for $M_1$ and $M_2$ that also scales with the Sun-spacecraft distance and search for local maxima.
We apply a Gaussian-convolution filter to the measured EMF with a small smoothing value of $\sigma_0=0.9$ to make the search for the maxima easier.
\pagebreak


\section{Choice of Parameters} \label{parameters}

\subsection{Event Characteristics} \label{events}

We expect that CMEs are not the only structures in the inner heliosphere to give rise to an EMF.
Turbulent solar wind streams or some magnetic transient events may also feature vortical plasma flows, helical magnetic field structures, and significant amplitudes of the EMF.
CMEs feature shocks in the magnetic field $B$, the plasma flow velocity $U$, the proton number density $n_p$ and the proton temperature $T_p$
Additionally, we expect strong peaks in the EMF $M_1$ and $M_2$ in the turbulent sheath region immediately after the shock front.
Other transient events with turbulent structures may not feature shocks in some of $B$, $U$, $n_p$ and $T_p$, but could still give rise to an EMF.
For the event list, we try to separate between CMEs and other transient events and introduce two categories:
CMEs are recorded into the {\em C}-type (``CME") category and other transient events with strong EMF measurements are put into the {\em T}-type (``Transient") category.
There might be some overlap between these two categories due to errors in the detection algorithm or weaknesses in the choice of the detection criteria.
Some weaker CMEs might erroneously be detected as {\em T}-types while some of the stronger non-CME events might be detected as {\em C}-types.

\subsection{Detection Criteria} \label{criteria}

To decide whether an event is recorded or not we introduce several criteria within an interval of $\pm k/2$ points.
The main criterion is on the EMF.
If there is no measurement in $M_1$ and $M_2$, no event is recorded.
Since $M_1$ is an unmodified formulation and was shown to be more consistent between model and observational results in~\cite{Bourdin+al:2018}, we use it as the stronger criterion.
The next most important parameter is the plasma flow velocity $U$ that should always peak around a CME shock front.
The detection criteria for the two categories; see Sec.\,\ref{events}; are given in Tab.~\ref{tab:criteria}.
Each box contains a set of criteria to be met for the respective event type.
If an event satisfies the criteria in at least one of these options, the event is recorded.
Events that meet the {\em C}-type criteria cannot be recorded a second time as {\em T}-types.

The {\em C}-type criteria are more strict and need to be triggered in at least three of the four quantities $U$, $B$, $n_p$ or $T_p$, as well as in the EMF.
{\em T}-type events must satisfy less strict criteria and may be recorded with only two triggers out of four if also a strong EMF is measured.
Note that the detection algorithm may produce some false shock front detections or miss a few of the shock fronts.
Therefore, it is possible that events are missed or recorded in the wrong category.

\begin{table}[ht]
\scriptsize
\setlength{\tabcolsep}{3.9pt}
\begin{center}
\caption{Shock Front Detection Criteria}
\label{tab:criteria}
\begin{tabular}{lcc}
    \hline
    \hline
    & {\em C}-type & {\em T}-type \\\noalign{\smallskip}
    \hline\noalign{\smallskip}
        & $M_1$ plus $M_2$  & $M_1$ plus $M_2$ \\
	a)  & trigger in $U$    & -- \\
	    & $\ge$ 2 triggers from $B$, $n_p$ or $T_p$ & $\ge$ 2 triggers from $U$, $B$, $n_p$ or $T_p$ \\\noalign{\smallskip}
	\hline\noalign{\smallskip}
	    & $M_1$     & $M_1$ \\
	b)  & --        & -- \\
	    & trigger in all of $U$, $B$, $n_p$ or $T_p$ & $\ge$ 3 triggers from $U$, $B$, $n_p$ or $T_p$ \\\noalign{\smallskip}
	\hline\noalign{\smallskip}
	    &      & $M_1$ \\
	c)  & --   & trigger in $U$ \\
	    &      & $\ge$ 1 trigger from $B$, $n_p$ or $T_p$ \\\noalign{\smallskip}
	\hline\noalign{\smallskip}
	    &       & $M_2$ \\
	d)  & --    & trigger in $U$ \\
	    &       & $\ge$ 2 triggers from $B$, $n_p$ or $T_p$ \\\noalign{\smallskip}
	\hline
\end{tabular}
\tablecomments{{\em C}-type (CME): strict criteria, higher chance of being a CME.
{\em T}-type (Transient): less strict criteria, smaller chance of being a CME. }
\end{center}
\end{table}

\subsection{Thresholds} \label{threshold}

\begin{table}[ht] 
\scriptsize
\setlength{\tabcolsep}{14.5pt}
\begin{center}
\caption{Thresholds at 1 au Used for Shock Front Recording and Scaling with the Sun-Spacecraft Distance $r_{S}$ Given in [au] }
\label{tab:threshold}
\begin{tabular}{lccr}
    \hline
    \hline
    Quantity & Symbol & Value & Scaling   \\ \noalign{\smallskip}
    \hline\noalign{\smallskip}
    $M_1$	& $\chi_{\rm M1}$ & $9 \times 10^{-5}~{\rm V~m^{-1}}$ & $r_{S}^{-1.5}$ \\
    $M_2$	& $\chi_{\rm M2}$ & $3 \times 10^{-5}~{\rm V~m^{-1}}$ & $r_{S}^{-1.5}$ \\
    \\
    $U$	& $\chi_{U}$ 	& $14~{\rm km~s^{-1}}$ 	& $r_{S}^{0}$ \\
    $B$	& $\chi_{B}$ 	& $0.8~{\rm nT}$	& $r_{S}^{-1.5}$ \\
    $n_P$	& $\chi_{N}$ 	& $1.5~{\rm cm^{-3}}$ 	& $r_{S}^{-2}$ \\
    $T_p$	& $\chi_{T}$ 	& $2.75 \times 10^{4}~{\rm K}$ & $r_{S}^{-1}$ \\\noalign{\smallskip}
    \hline
\end{tabular}
\end{center}
\end{table}

The detection thresholds for the EMF and the shock fronts and their scaling with the Sun-spacecraft distance $r_{S}$ in [au] are given in Tab.~\ref{tab:threshold}.
The sampling rate is $\Delta t = 648$~s and the width of $\overline{P_n}$ is set to $k = 12 \cdot \Delta t = 2.16$~hr.
Keep in mind that the thresholds of $U$, $B$, $n_p$ and $T_p$ are applied to the derivative of the moving median-difference $\frac{\partial}{\partial t}\overline{P_n}$.
With respect to the heliocentric distance we use a threshold for the absolute value of the plasma bulk velocity \citep{Khabarova+al:2018}, the magnetic field magnitude \citep{Behannon:1978}, the proton density \citep{Eyni-Steinitz:1980}, and the proton temperature \citep{Lamarche+al:2014}.


\section{Results} \label{results}

\subsection{Event List} \label{list}

\begin{figure}[ht]
\begin{center}
\includegraphics[trim=0 0 0 0,width=8.0cm]{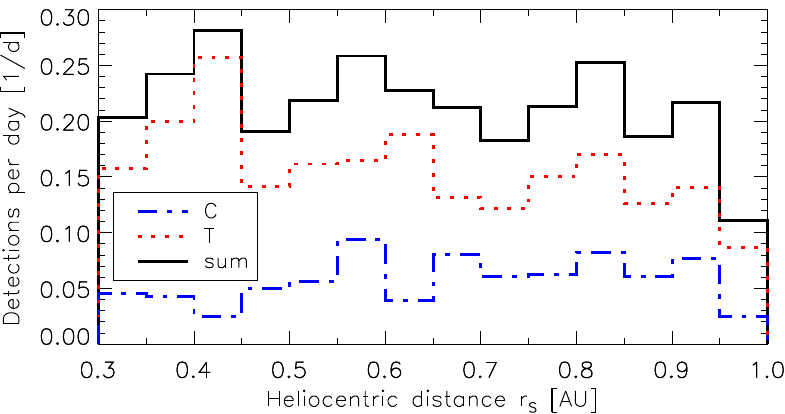}
\end{center}
\vspace*{-5mm}
\caption{Recorded shock fronts normalized by the observation time with heliocentric distance.
Combined {\em Helios-1} and {\em Helios-2} events (531 total) have been used.}
\label{fig:hist_norm}
\end{figure}

For the {\em Helios-2} spacecraft between 1976 January 17 and 1980 March 8, we find 176 shock fronts with 46 of category {\em C} and 130 of category {\em T}.
There is no detection after 1979 May 16 because there are no magnetic field measurements afterwards.
The {\em Helios-1} observations range from 1974 December 12 to 1985 November 4 and the magnetic field measurements are deactivated on 1981 June 25.
We obtain 355 shock fronts of which 99 are in category {\em C} and 256 are in category {\em T}.
The detection rate per day is shown in Fig.~\ref{fig:hist_norm}.
We exclude intervals from the calculation where one or all of the instruments were shut down.

On average, we find one event every 4-5 days at most heliocentric distances.
Closer to Earth's orbit, the detection rate seems to decrease to a minimum of one event every $9$ days between $0.95$ and $1$~au.
The CME-like events ({\em C}-type) occur about every 15-20 days with no significant dependence on the heliocentric distance.
\begin{figure}[ht]
\begin{center}
\includegraphics[trim=0 0 0 0,width=8.0cm]{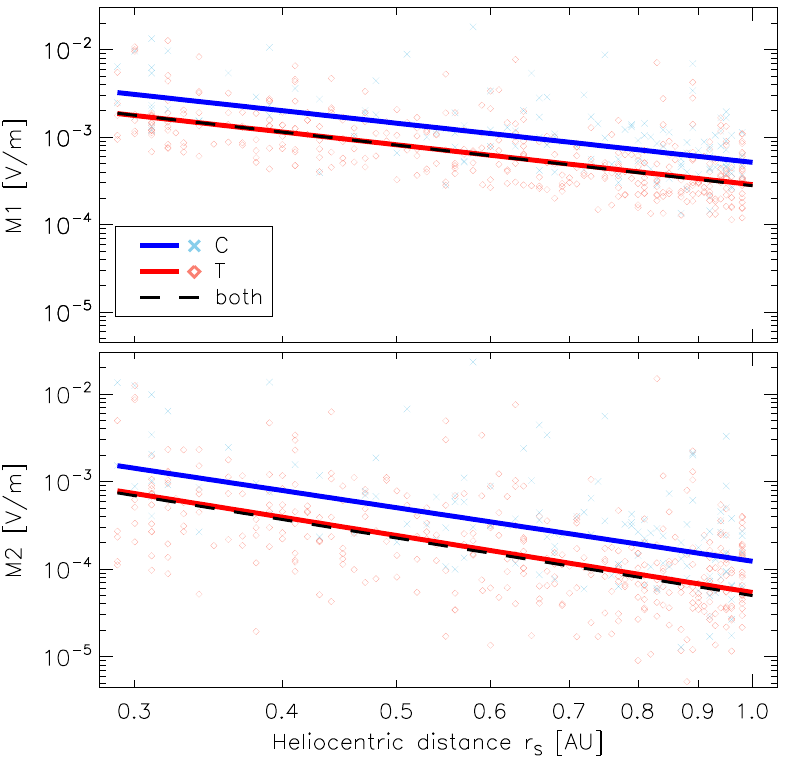}
\end{center}
\vspace*{-5mm}
\caption{
Dependence of the EMF at the recorded shock fronts on the heliocentric distance.
Combined {\em Helios-1} and {\em Helios-2} events (531 total) have been used.
Power-law dependencies are given for {\em C} and {\em T} events separately, as well as a combined result.
Upper panel: $M_1$ from Eq.~(\ref{eq:em}): {\em C} $\propto r_S^{-1.48}$, {\em T} $\propto r_S^{-1.51}$, both $\propto r_S^{-1.54}$.
Lower panel: $M_2$ from Eq.~(\ref{eq:em2}): {\em C} $\propto r_S^{-2.03}$, {\em T} $\propto r_S^{-2.13}$, both $\propto r_S^{-2.18}$.
}
\label{fig:scatter}
\end{figure}
For the {\em T}-type events we get an average detection rate of one event per 7 days. 
We see a notable descent in detections from small to large heliocentric distances.

Our results agree well with the assumptions we use for our detection criteria. 
CME detections remain roughly constant at around two detections per month, compared to two to three detections per month as found by~\cite{Richardson-Cane:2010}.
On the other hand, we assumed that the {\em T} events consist of non-CME events with a strong enough turbulence to feature a significant EMF.
{\em T} events may include, for example, turbulent fast solar wind streams or corotating interaction regions.
Therefore, it is not surprising that these structures would become less turbulent at larger heliocentric distances and are then no longer detected by our criteria.

\subsection{Power-law Fit} \label{power}

We analyze the EMF in relation to the heliocentric distance; see Fig.~\ref{fig:scatter}.
For the recorded events, we fit a power law as a linear regression line in the double-logarithmic data with a least absolute deviation method.
This method is more robust to outliers than a least-squares minimization.
We calculate three separate power laws for $M_1$ and $M_2$:
one from the {\em C} events $E_C$, one from the {\em T} events $E_T$ and one from the combined events $E_{{\rm both}}$.
However, we want to avoid that the {\em T} events flatten out the slope of the regression line because of their commonly lower EMF measurements compared to the {\em C} events.
Therefore, we lower the values of the {\em C} events to match the average of the {\em T} events. 

We find the exponents $E_C(M_1)=-1.48 \pm 0.03$ and $E_T(M_1)=-1.51 \pm 0.03$ for the {\em C} and {\em T} events in $M_1$.
The exponent of the combined events is $E_{{\rm both}}(M_1)=-1.54 \pm 0.02$, which is close to the scaling we applied in our calculation of the event list from Tab.~\ref{tab:threshold} ($\sim r_S^{-1.5}$).
The exponents of $M_2$ are $E_C(M_2)=-2.03 \pm 0.03$, $E_T(M_2)=-2.16 \pm 0.03$, and $E_{{\rm both}}(M_2)=-2.18 \pm 0.02$, which is larger than the estimated scaling from Tab.~\ref{tab:threshold}.
A possible explanation for this inconsistency could be the method we use to calculate $M_2$ in comparison to $M_1$.
While $M_1$ is calculated directly from the cross product of $\delta U$ and $\delta B$, $M_2$ also contains the curls of these quantities; see Eq.~(\ref{eq:em}) and (\ref{eq:em2}).
Because \cite{Bourdin+al:2018} use single-spacecraft data, these authors apply a method that replaces the spatial derivatives with the time derivative in the direction of the mean solar-wind flow. 
This might lead to a stronger dependence of the radial magnetic field component, which is known to scale proportionally to $\sim r_S^{-2}$, while only the magnitude of the magnetic field is proportional to $\sim r_S^{-1.5}$ \citep{Behannon:1978}.

\subsection{Comparison to Other Works} \label{ipshocks}

We compare our event list to the {\em IPshocks}\footnote[3]{\url{http://ipshocks.fi/}\label{link:ips}} database of the {\em University of Helsinki} used by \cite{Kilpua+al:2015}.
The events in the {\em IPshocks} database are found by visual inspection of daily plots of magnetic field and plasma parameters.
The preselected shock candidates then need to fulfill the following fast-forward (FF) shock criteria:
(1) $B_{{\rm down}}/B_{{\rm up}} > 1.2$, (2) $n_{{\rm down}}/n_{{\rm up}} > 1.2$, (3) $T_{{\rm down}}/T_{{\rm up}} > 0.83$, (4) $U_{{\rm down}}-U_{{\rm up}} > 25$ km~s$^{-1}$, and (5) the upstream magnetosonic Mach number $M_{{\rm ms}}>1$. 

For the {\em Helios} spacecraft, \cite{Kilpua+al:2015} find a total of 102 FF shocks.
We reproduce 73 of their events ($71.6\%$) when we allow a time difference of up to 2.5 hr.
46 ($63.0\%$) of these events are detected as {\em C}-type, while the other 27 ($37.0\%$) are recognized in the {\em T}-type category.
Despite the fundamental differences in the detection methods, our fully automated method proves to be able to detect the majority of the interplanetary CMEs.
We are confident that the agreement between both lists can be further improved by the application of more sophisticated detection criteria and thresholds to our method.

\begin{figure}[ht]
\begin{center}
\includegraphics[trim=0 0 0 0,width=8.0cm]{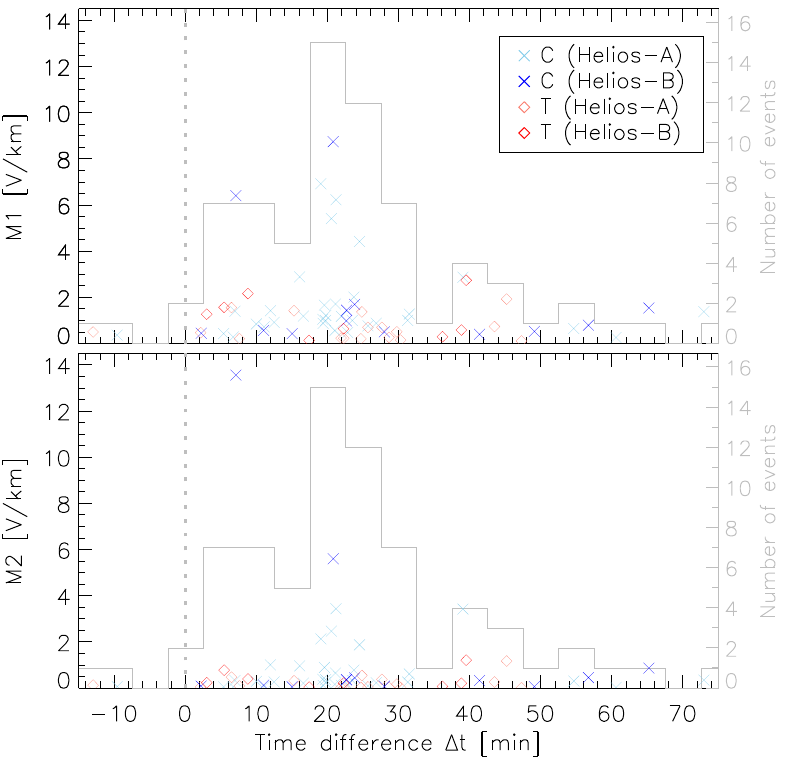}
\end{center}
\vspace*{-5mm}
\caption{
Comparison of the arrival times of the detected C and T events to the 73 shocks from \cite{Kilpua+al:2015}.
We plot the time difference against the EMF peak value.
The histogram gives the amount of events within each bin. 
Upper panel: $M_1$ from Eq.~(\ref{eq:em}).
Lower panel: $M_2$ from Eq.~(\ref{eq:em2}).
}
\label{fig:td}
\end{figure}

The time difference of the C and T events to the FF shocks in the {\em IPshocks} database is plotted in Fig.~\ref{fig:td}.
We need to note that our automated detection method records the events at the peak of the measured EMF.
We observe that the majority of events in our list are recorded between 0 and 50 minutes after the shocks in the {\em IPshocks} database.
The strongest amplitudes in the EMF are measured around 20 minutes after the shock front.
As expected, the EMF rises significantly in the sheath regions following the shock fronts of the detected ICMEs and magnetic transient events.

\pagebreak


\section{Discussion} \label{discussion}


We follow the method of~\cite{Bourdin+al:2018} where we apply EMF as a shock front indicator and create a list of magnetic transient events from the {\em Helios} observation using the EMF as a central detection criteria. 
Additionally, we apply a shock-front detection algorithm that uses a moving median-difference to search for shock fronts in the magnitude of the magnetic field $B$, the plasma flow velocity $U$, the proton number density $n_p$, and the proton temperature $T_p$.

With our automated detection method, we find a constant detection rate of about two CME-like events per month at all heliocentric distances.
This result fits well to the two to three detections per month found by~\cite{Richardson-Cane:2010}.
At the same time, we find a noticeable increase in strong EMF measurements at small heliocentric distances, which leads to an increased detection rate in the second event category of non-CME transients.

We fit a power law for the peak EMF magnitude versus the heliocentric distance $r_S$.
For the mean-field formulation $M_1$ we find a dependence of $r_S^{-1.54}$ close to the value of $~ r_S^{-1.5}$ that we anticipated.
The adapted formulation of $M_2$ from P.-A. Bourdin (2018, private communication) shows a dependence of $r_S^{-2.18}$ that is closer to the expected scaling of the radial magnetic field ($~ r_S^{-2}$).
This can be understood by taking a look at the calculation method of the curls of the magnetic field $\delta B$ and plasma flow velocity $\delta U$ fluctuations.
Due to the limitations of single-spacecraft data, the spatial derivative has to be replaced with the time derivative in the mean solar-wind direction, which might explain the scaling similar to the radial field component.

Our event list comprises most of the events from the {\em IPshocks} database \citep{Kilpua+al:2015} and also 99 additional CME-like events.
We reproduce $71.6\%$ of the manually identified shocks with our automated detection method.
When we look at the time difference of the arrival times between the two databases, we find that most events from our list are detected from 0 to 50 minutes after the {\em IPshocks} table lists the shock front.
Since our method sets the arrival time at the center of the EMF peak, this suggests that the EMF measurements are strongest in the turbulent sheath region immediately following the shock front.

Especially near Earth's orbit, the EMF appears to be a good indicator for the arrival of shock fronts as there are significant peaks at most of the transient events while the overall fluctuations are small.
Closer to the Sun, we find some highly turbulent regions with strong fluctuations in the EMF that are of similar amplitude as generated in the sheath regions of actual shock fronts.
This can lead to increased false detections or missing of weaker shock fronts, so the scaling and the choice of the EMF thresholds have to be as accurate as possible.
The thresholds of the magnetic field and plasma parameters are equally important to avoid false detections from the median-difference algorithm.
However, prominent shock fronts are detected fairly well by the method and we are confident that the detection method can be further improved by other choices for the thresholds and their scaling, choosing the width of the moving median-difference $k$, or applying different detection criteria.

Checking daily plots of the magnetic field and plasma measurements, we occasionally find shocks that do not feature a prominent peak in both $M_1$ and $M_2$ where an immediately preceding shock front was detected.
A possible explanation could be that the preceding shock front pushes the solar-wind material out of the way so that there is not much interaction at the second shock front. 
At closer distances to the Sun, there are cases where we find an enhancement of the EMF while there is no clear shock front in the plasma flow velocity, the magnetic field, the number density, and the proton temperature. 
Those regions could be highly turbulent fast solar-wind streams or corotating interaction regions and might be interesting to study more closely in future works.
Because our method works well at distances close to Earth's orbit, one could apply our method to data from more modern space missions and compare them with visual images.

\subsection{Outlook} \label{outlook}

In this work, we found the EMF to be a good indicator of turbulent peaks of CMEs and magnetic transient events in the solar wind.
The EMF is an easy to calculate scalar quantity that could be used on board spacecraft to automatically decide whether to switch to the highest sampling rate when a turbulent structure passes by without the need to observe multiple observational quantities or manual interaction.
The quality of our detection method could be improved by redefining the thresholds and the scaling with the heliocentric distance.
For example, instead of setting a fixed threshold with a scaling, one could dynamically calculate a base-level of each quantity in each individual data interval.
Such a dynamic scaling would improve the identification of significant peaks and avoid false detections. 
An interesting idea for future work is to perform a superposed-epoch analysis of the combined {\em C} or {\em T} events (P.-A. Bourdin 2018, private communication).
With this method one superposes the observational data of all the extracted individual events to a combined average signal per category.
The results of the analysis will give another hint to whether the event categories and the detection criteria in our study are chosen well.

The observation that the EMF peaks on average 20 minutes after the initial shock front supports the idea of~\cite{Bourdin+al:2018} that the EMF can be used
for studies of the inner structure of CMEs.
However, such a study will likely need to investigate the magnetic field and plasma velocity components at a higher time resolution than the {\em Helios} spacecraft provide.
{\em Solar Orbiter} is an upcoming spacecraft mission to be launched in 2020 that will orbit the Sun at a similar heliocentric distance as {\em Helios} and will provide a higher time resolution of the in situ data, as well as remote-sensing measurements for visual images of the transient events.
Alternatively, there are plenty of spacecraft orbiting near $1$ au that could provide data for CME analyses.

Future work may include an extended statistical study of the EMF during different periods of the solar cycle.
Additionally, one should compare measurements in the slow solar wind with measurements in the fast solar wind, as well as measurements of different structures such as stream interaction regions.


\section*{} \label{acknowledgments}

This work is financially supported by the Austrian Space Applications Programme at the Austrian Research Promotion Agency, FFG ASAP-12 SOPHIE under contract 853994.
The observational data is provided by the {\em Helios} data archive in CDAWeb\footnote[4]{\url{https://spdf.sci.gsfc.nasa.gov/pub/data/helios/}} located at the Space Physics Data Facility (NASA/GSFC).


\bibliography{lit}
\bibliographystyle{aasjournal}

\end{document}